# Unconventional Spin-Glass-Like State in AgCo$_2$V$_3$O$_{10}$, the Novel Magnetically Frustrated Material

M. Hadouchi[a,*], A. Assani[a], M. Saadi[a], Y. Kopelevich[b,*], R. R. da Silva[b], A. Lahmar[c], H. Bouyanfif[c], M. El Marssi[c], L. El Ammari[a]

[a]Laboratoire de Chimie Appliquée des Matériaux, Centre des Sciences des Matériaux, Faculty of Sciences, Mohammed V University in Rabat, Avenue Ibn Battouta, BP 1014, Rabat, Morocco.

[b]Instituto de Física "Gleb Wataghin", Universidade Estadual de Campinas, Unicamp 13083-859, Campinas, São Paulo, Brasil.

[c]Laboratoire de Physique de La Matière Condensée (LPMC), Université de Picardie Jules Verne, Amiens, France.

*Corresponding authors:

Hadouchi Mohammed; E-mail: hadouchimohammed@yahoo.com

Yakov Kopelevich: E-mail: kopel@ifi.unicamp.br

**Abstract**

Single crystals of a new silver and-cobalt based vanadate AgCo$_2$V$_3$O$_{10}$ were grown from a melted mixture. The crystal structure determination reveals that this new vanadate crystallizes in triclinic system with space group $P\bar{1}$. The structure of the titled compound is constructed from CoO$_6$ octahedra and VO$_4$ tetrahedra sharing edges and vertices leading to an open three-dimensional framework delimiting tunnels along [001], where the Ag cations are located. The bands observed in Raman spectrum were assigned to corresponding vibrations of the VO$_4$ groups. DC and AC magnetization (susceptibility) measurements revealed the spin-glass (SG) -type transition below a frequency-independent temperature $T_f(H)$ that is nearly independent on the applied magnetic field. Magnetic field $H > 10$ kOe induces the spin glass- antiferromagnet. These observations brings AgCo$_2$V$_3$O$_{10}$ into the class of geometrically frustrated magnetic systems.

# 1. INTRODUCTION

Nowadays, physics of spin glasses (SG), studied during the last several decades [for a review article see K. Binder, A.P. Young, Rev. Mod. Phys. 58, 801 (1986)[1]], gains the interdisciplinary interest. It has been widely believed that randomness and structural defects are necessary for the glassy behavior. However, recent studies provided both experimental and theoretical evidence that SG-like state can occur in frustrated magnetic systems for arbitrary weak quenched disorder. Thus, spin freezing has been analyzed for geometrically frustrated antiferromagnets on the pyrochlore lattice and the occurrence of the SG-like transition has been anticipated for many geometrically frustrated magnets [2]. Exotic spin ice and spin liquid topologically ordered ground states have been also found[3,4]in geometrically frustrated magnets. The behavior different from that known for conventional SG has been reported in those systems, such as e.g.(1) magnetic-field- driven spin glass – spin liquid – antiferromagnet transition [5];(2) the freezing temperature $T_f(H)$ enhancement with the applied magnetic field and its insensitivity to frequency of applied ac field [6,7]; (3) jaming phenomenon [8] resulting in a step-like non-equilibrium time relaxation [9], known for structural glasses.

Because of the geometrical frustration and strong magnetic anisotropy, a considerable interest to magnetic properties of cobalt-based materials has also been arisen[4,10,11]. Among the cobalt-based compounds that have been magnetically characterized, one can mention as examples, $Ca_3Co_2O_6$ [12,13], which was reported exhibiting a field-induced transition from ferrimagnetic to ferromagnetic structure and strong magnetic anisotropy, whereas $Co_3O_4$ nanocrystallites[14], reported showing interference effects between superparamagnetic and spin glass correlated moments. In addition, a spin-glass-like dynamic relaxation is observed in the cobalt-based hydroxysulfate, i.e, $Na_2Co_3(OH)_2(SO_4)_3(H_2O)_4$ [15] exhibiting ferrimagnetic chains, while the authors Vilminot et al.[16], reported a large coercivity of 70 kOe at 1.8 K in the ferrimagnet, $K_2Co_3(OH)_2(SO_4)_3(H_2O)_2$. Moreover, several inorganic compounds and cobalt-organic frameworks with spin-cluster and/or spin-chain structures have been studied[17–23]. For instance, Cobalt-based compounds showing one-dimensional spin-chain structure such as $Co_3(BPO_4)_2(PO_4)(OH)_3$[19]and $CoCl_2.2H_2O$ [24] were reported with a step-like magnetization curves. Furthermore, the cobalt-containing garnet, $CaY_2Co_2Ge_3O_{12}$ [25], was reported with unusual anisotropic and chainlike antiferromagnetism.

In fact, only some cobalt-based vanadates that have been magnetically investigated. For instance, we can mention,$CoV_2O_6$ [26] with a spin-chain structure showing large anisotropy and 1/3 magnetization plateau, while $BaCo_2V_2O_8$ [27]exhibits a quantum critical point.

In the context of synthesizing new magnetic materials, we report in this work the synthesis and characterizations of a novel cobalt-based vanadate $AgCo_2V_3O_{10}$. Our studies revealed magnetic properties of this new material, which are similar to those reported for geometrically frustrated magnetic systems, introduced above.

## 2. EXPERIMENTAL SECTION

### 2.1. Synthesis

#### 2.1.1. Single crystal synthesis

Crystals of $AgCo_2V_3O_{10}$ were obtained during the exploration of $A_2O/CoO/Cr_2O_3/V_2O_5$ quaternary system. The reactants, $AgNO_3$ (≥ 99%, Acros Organics), Cobalt metal powder (99.88 %, Scharlau), $Cr(NO_3)_3 \cdot 9H_2O$ (≥ 98.0 %, Merck) and $V_2O_5$ (≥ 99.6 %, Acros Organics) were mixed in a proportion corresponding to the molar ratio of 2 : 2 : 1 : 3/2 respectively. The mixture was placed in a platinum crucible and then heated gradually until melting point of 1153 K and held for 5 hours. Single crystals were grown by cooling the molten mixture to room temperature at rate of 5Kh$^{-1}$. The resulting mixture contained black crystals with a suitable size for the X-ray diffraction measurements.

#### 2.1.2. Powder synthesis

The powder sample was obtained by sol-gel method. In a first step a stoichiometric proportion corresponding to desired composition of the reactants $AgNO_3$ (≥ 99%, Acros Organics), $(CH_3COO)_2Co \cdot 4H_2O$ (≥ 99.0 %, Merck) and $V_2O_5$ (≥ 99.6 %, Acros Organics) was dissolved in appropriate amount of distilled water and a few drops of $HNO_3$ and kept under stirring. Secondly, citric acid (≥ 99.5 %, Merck) was added to the mixture in a 1:1 molar ratio. The solution was evaporated slowly to form a viscous liquid that was kept under heating to dryness. The resulting gel was transferred to furnace to undergo successive heat treatments with intermittent grinding until 813 K. A black powder was obtained and the purity was confirmed by powder X-ray diffraction.

### 2.2. Single crystal X-ray Diffraction

A black crystal with appropriate size (0.32× 0.24 × 0.19mm$^3$) was selected for single-crystal data collection at room temperature on a Bruker X8 Apex diffractometer equipped with an Apex II CCD detector with Mo Kα radiation, λ= 0.71073 Å and Graphite monochromator. The software APEX2 was used for data collection and SAINT for cell refinement and data reduction. A total number of 24359reflections were measured in the range of $\theta_{min} = 2.3°$ and $\theta_{max} = 36.1°$, of which3831were independent and 3480 reflections with I >2σ(I). The crystal structure was solved using direct method and refined by SHELXS 2013[28] and SHELXL 2013[29]programs incorporated in the WinGX[30] program. Absorption corrections were performed from equivalent reflections based on multiscans. The atomic displacement parameters for all atoms were refined anisotropically. For molecular drawing, the programs ORTEP-3[30] and DIAMOND[31] were used.

## 2.3 Additional characterizations

The morphology and elemental analysis of the synthesized single crystal and powder were performed with JEOL JSM-IT100 InTouchScope™ scanning electron microscope equipped with energy dispersive X-ray spectroscopy analyzer (EDS).

To control the purity of the synthesized powder, an X-ray powder diffraction pattern was obtained at room temperature using a Siemens D5000 powder diffractometer operating with θ-2θ scan mode and CuKα radiation (λ = 1.5406 Å). The data were collected over the 2θ angle range of 10° ≤ 2θ ≤70° with a step size of 0.03° and 30s per step counting time.

Raman spectroscopy data were collected using a Renishaw in Via Qontor Raman microscope on powder sample with 532 nm laser as excitation wavelength. Laser power has been optimized (0.5 mW) to avoid overheating of the sample. The spectrum was recorded in backscattering geometry from 50 to 1300 cm$^{-1}$.

Magnetization measurements were performed by means of a Physical Property Measurement System (PPMS) DynaCool magnetometer and SQUID magnetometer (Quantum Design) on the synthesized powder (15.1 mg - Sample 1, and 33 mg – Sample 2) sealed in a gelatin capsule. Our synthesis method has allowed isolating single crystals with a rather smaller size, not sufficient to carry out measurements on individual crystals.

The isothermal magnetization curves $M(H)$ were measured up to4T. The temperature dependencies of magnetization $M(T)$ were recorded in the temperature interval 2 K <$T$< 300K and an applied magnetic field up to $H$ = 4 T in both ZFC (zero-field-cooled) and FCC (field cooled on cooling) regimes. AC susceptibility measurements in the temperature range of 2-60 K were performed with AC field of 10 Oe and frequencies between 500 Hz and 5 kHz. We have also done magnetization time relaxation measurements $M(t)$ under ZFC conditions. In all the data, the diamagnetic signal of all atoms (-184·10$^{-6}$ emu.mol$^{-1}$) has been subtracted.

## 3. RESULTS AND DISCUSSION

### 3.1. Crystal structure determination and description

This new vanadate crystallizes in triclinic system with space group $P\bar{1}$. The crystal structure of this vanadate is characterized by a disorder in the Ag$^+$ position, which is splitted in two positions. The refinements of the occupancy factors of silver atoms leading to values of Ag1: Ag2 = 0.51 (1): 0.49 (1), which led to the exact formula of the title compound. As a matter of fact, the structure of this vanadate is isotype to NaMg$_2$V$_3$O$_{10}$[32], except the Ag disorder. The linkage of two CoO$_6$octahedra, threeVO$_4$tetrahedraandtwo Ag$^+$ cations forming the asymmetric unit of AgCo$_2$V$_3$O$_{10,}$ is showninFigure.1. In this structure all atoms are in the general position 2i,and the atoms Co1 and Co2 have octahedral environments forming Co$_4$O$_{18}$ tetramericunits through edge-sharing (Figure. 2a).The distances between Co2─Co1 and Co1─Co1 are 3.146(2)Å and 3.210(2)Å respectively (Figure. 2b). Vanadium atoms adopt tetrahedral environments VO$_4$which three of these tetrahedral share corners to form V$_3$O$_{10}$ groups. In this structure, the succession of the Co$_4$O$_{18}$ units through corner- sharing with two V$_3$O$_{10}$ groups leads to infinite ribbons along [010], (see Figure.3). The stacking of these ribbons defines an open three-dimension framework delimiting tunnels parallel to [001] where the Ag$^+$ cations are located (Figure.4). In the structure of this vanadate all sites are fully occupied excluding Ag sites. The atom Ag1 is coordinated by seven oxygen atoms, with Ag1─O distances in a range between 2.347 (2) and 3.229 (2) Å. While Ag2 is surrounded by eight oxygen atoms with Ag2─O distances in a range between 2.354(3) and3.314(8) Å. Bond valence sum calculations of all atoms (Brown & Altermatt, 1985)[33] in the structure are in reasonable agreement with the oxidation states. Bond valence sums (in valence units) for AgCo$_2$V$_3$O$_{10}$ are 0.95 for Ag1, 0.896 for Ag2, 2.13 for Co1, 2.03 for Co2, 5.0 for V1, 4.91 for V2 and 4.88 for V3. Noting that the values of the bond valence sums calculated for all oxygen atoms are between 1.92 and 2.16.Crystal data, data collection and structure refinement details are summarized in Table 1. Fractional atomic coordinates, atomic displacement parameters and selected bond lengths and angles are given in Table S1, Table S2 and Table S3 (in supplementary material).

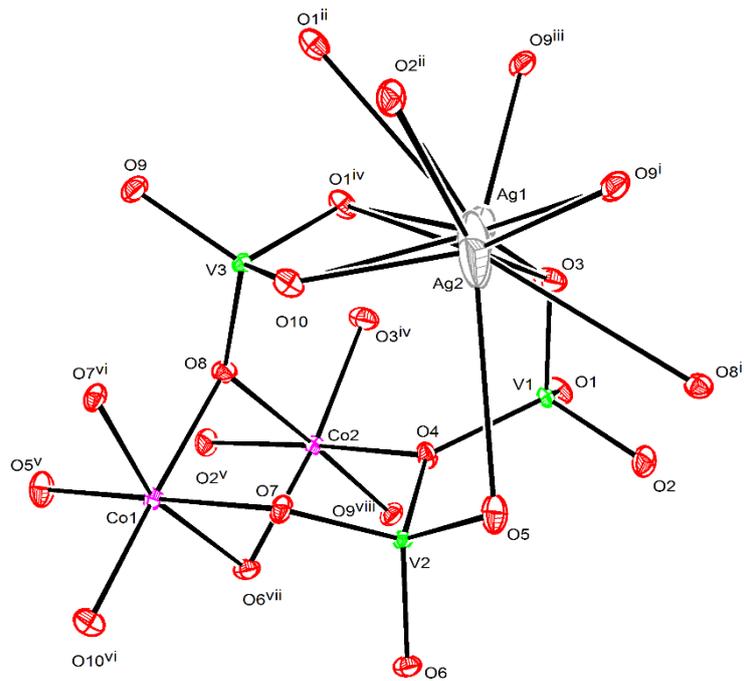

Figure1. ORTEP plot of the principal building units in the structure of AgCo$_2$V$_3$O$_{10}$. Displacementellipsoids are drawn with 50% probability level. Symmetry codes: (i) *x*+1, *y*, *z*; (ii) *x*, *y*+1, *z*; (iii) -*x*+1, -*y*+1, -*z*+1; (iv) -*x*+1, -*y*, -*z*+1; (v) *x*-1, *y*, *z*; (vi) -*x*+1, -*y*+1, -*z*+2; (vii) -*x*+1, -*y*, -*z*+2; (viii) *x*, *y*-1, *z*.

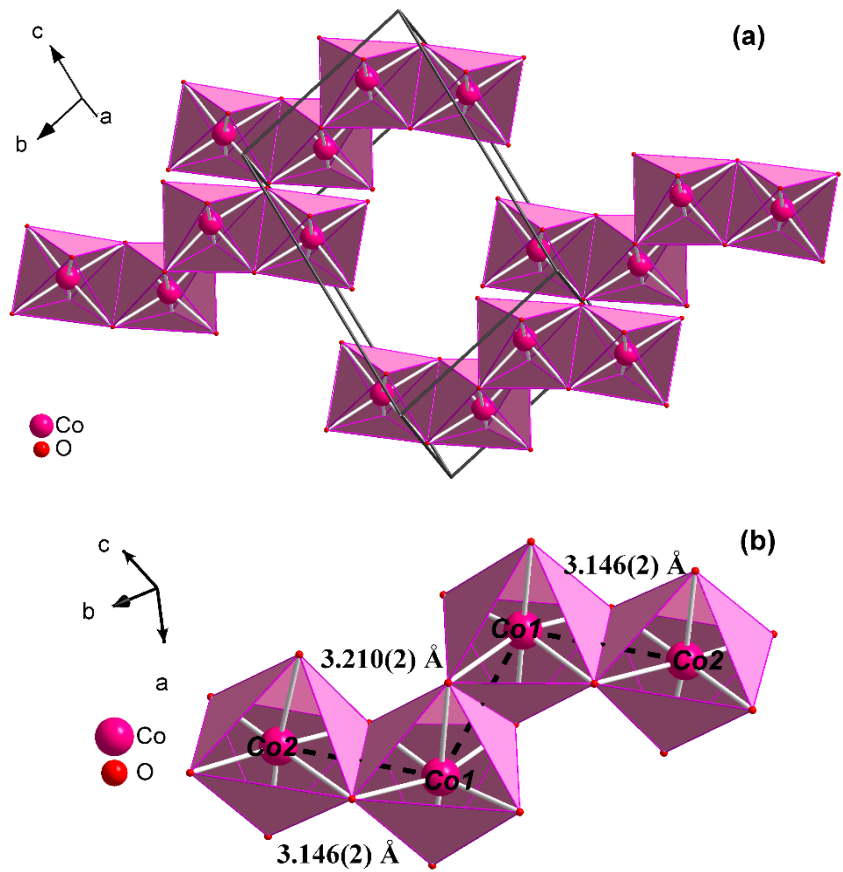

Figure 2. Polyhedral representation showing $Co_4O_{18}$ units (a) and (b) distances between the cobalt atoms forming the $Co_4O_{18}$ units.

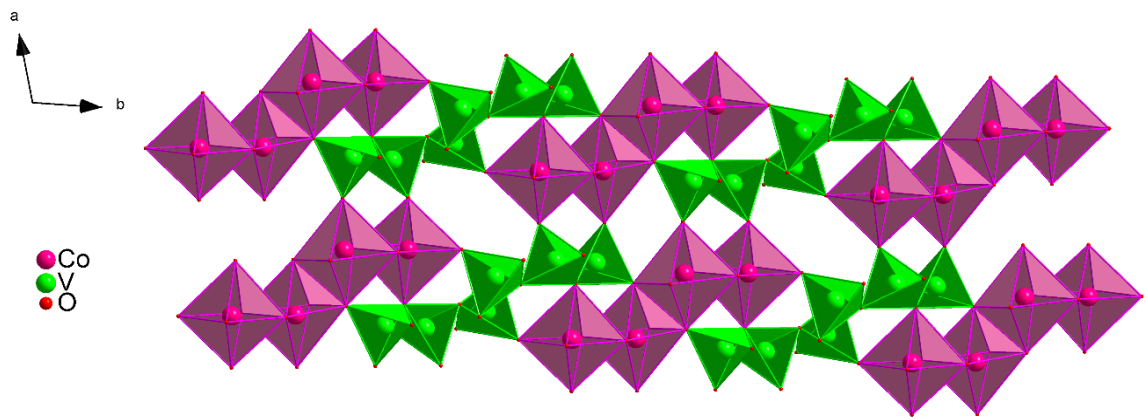

Figure 3. Sequence of $Co_4O_{18}$ units through corner- sharing with two $V_3O_{10}$ groups forming an infinite ribbon running along [010].

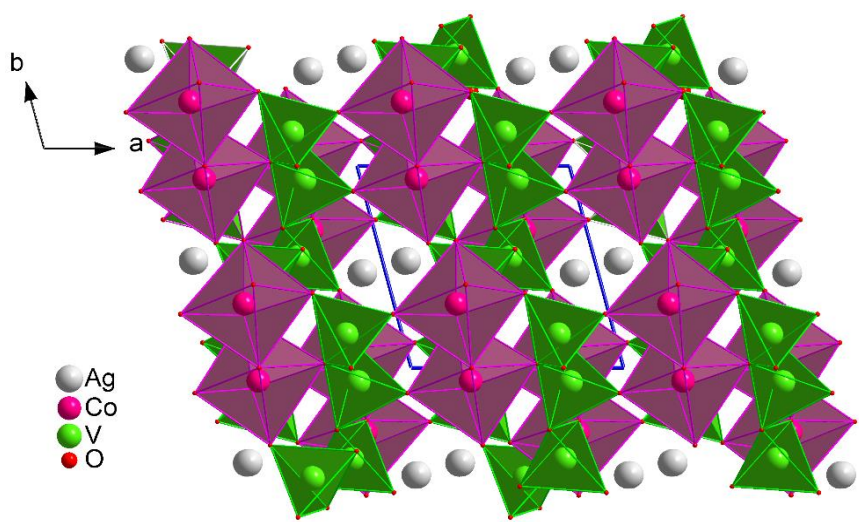

Figure 4. Polyhedral representation of $AgCo_2V_3O_{10}$ showing silver cations in the channels extending along [001].

Table.1. Crystal data, data collection and structure refinement details of $AgCo_2V_3O_{10}$.

| Crystal Data | |
|---|---|
| $AgCo_2V_3O_{10}$ | $\gamma = 101.803 (1)°$ |
| $M_r = 538.54$ | $V = 402.42 (1)$ Å$^3$ |
| Triclinic, $P\bar{1}$ | $Z = 2$ |
| $a = 6.7121 (1)$ Å | $F(000) = 500$ |
| $b = 6.7576 (1)$ Å | $Dx = 4.444$ Mg m$^{-3}$ |
| $c = 9.6057 (2)$ Å | Mo Kα radiation, $\lambda = 0.71073$ Å |
| $\alpha = 104.118 (1)°$ | $\mu = 9.79$ mm$^{-1}$ |
| $\beta = 99.748 (1)°$ | $T = 293$ K |
| **Data collection** | |
| 24359 measured reflections | $\theta_{max} = 36.1°$, $\theta_{min} = 2.3°$ |
| 3831 independent reflections | $h = -11 \rightarrow 11$ |
| 3480 reflections with $I > 2\sigma(I)$ | $k = -11 \rightarrow 11$ |
| $R_{int} = 0.036$ | $l = -15 \rightarrow 15$ |
| **Refinement** | |
| Refinement on $F^2$ | 0 restrictions |
| Least-squares matrix: full | $w = 1/[\sigma^2(F_o^2) + (0.0313P)^2 + 1.8197P]$ |
| $R[F^2 > 2\sigma(F^2)] = 0.027$ | where $P = (F_o^2 + 2F_c^2)/3$ |
| $wR(F^2) = 0.075$ | $(\Delta/\sigma)_{max} = 0.001$ |
| $S = 1.05$ | $\Delta\rangle_{max} = 1.07$ e Å$^{-3}$ |
| 3831 reflections | $\Delta\rangle_{min} = -2.89$ e Å$^{-3}$ |
| 156 L.S. parameters | Extinction correction: *SHELXL*, $Fc^* = kFc[1+0.001*Fc^2\lambda^3/\sin(2\theta)]^{-1/4}$ |
| | Extinction coefficient: 0.0025 (6) |

## 3.2. Powder X-ray Diffraction, SEM and EDS analysis.

The morphology and elemental analysis of the as-prepared single crystal and powder of $AgCo_2V_3O_{10}$ were characterized by scanning electron microscope (SEM) and energy dispersive X-ray spectrometer (EDS). The SEM images are presented in Figure.5 and revealed the formation of spherical particles with micrometric sizes in powder sample. The EDS analysis (Figure. S1 in supplementary material) confirms the presence of only Co, V, Ag and oxygen atoms, also Co/V and Ag/V ratios are really close to those of the elemental composition. Powder X-ray Diffraction confirmed the purity of the as-synthesized powder. The obtained X-ray pattern was fitted using Le Bail refinement method with JANA2006 software[34,35]. This refinement led to a good agreement between the experimental and the calculated patterns (see Figure.S2 in supplementary material).

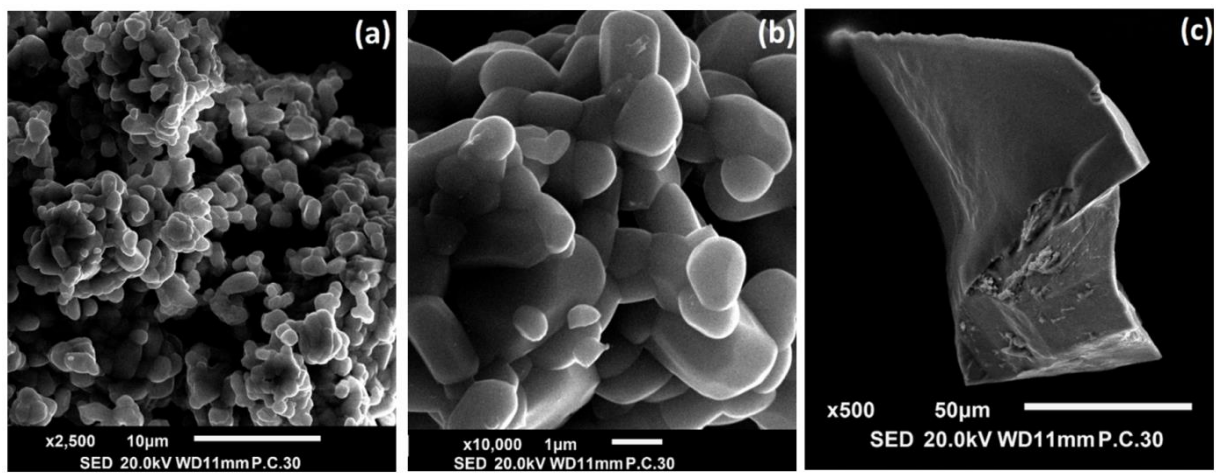

Figure 5. SEM micrographs showing the morphology of the as-synthesized powder (a), (b) and single crystal (c) of $AgCo_2V_3O_{10}$.

## 3.3. Raman spectroscopy

Figure 6, shows Raman spectrum obtained in the frequency range of 50-1300 cm$^{-1}$. Noting that the Raman investigation herein is recorded in polycrystalline form in backscattering configuration because the grown crystal has not a sufficient size to carry out polarized Raman investigation. Thus, the overlap of the vibrational modes is expected. Furthermore, it is well known that vanadates are affected not only by crystal symmetry but also by the presence of alkali, alkaline earth and transition metals. In order to give a reasonable assignment of the observed bands, we can divided the spectrum at 500 cm$^{-1}$. In the higher frequency range (above 500 cm$^{-1}$), the two bands at 554 and 662 cm$^{-1}$ could be assigned to the symmetric and antisymmetric stretching modes of bridging V─O─V units[36–38] corresponding to V2-O4-V1 and V1-O1-V3 with angle of 123.54 and 130.76 ° respectively. A closely frequencies are reported in some minerals vanadates , in these structures the frequency position is governed by the involved vanadium group units. However, the two sharp bands observed at 831 and 918 cm$^{-1}$ are assigned to the symmetric stretching of the shortest V─O bonds[39], namely V1—O1, V1—O2, V1—O3, V2—O5, V2—O7, V2—O6, V3—O8, V3—O9 and V3—O10 (Table S3). It is worth mentioning that the distance of these bands are very close which is favourable for the overlap of the vibrational modes, thus the observation of two modes.

Furthermore, the remaining bands located in 769 and 738 cm$^{-1}$ can be assigned to the stretching modes of the longer V—O bonds, i.e., V1—O4, V2—O4 and V3—O1$^{iv}$ (Table S3).

Concerning the second part of the spectrum below 500 cm-1, the identification of modes is not easy, especially between 200 and 450 cm$^{-1}$ with several overlapping bands. Nevertheless, in this region The V—O bending modes are typically ascribed [40]. Finally, the weak bands observed in the low wavenumber are probably due to the deformation VO subunits and MO bonds (M= Ag, Co).

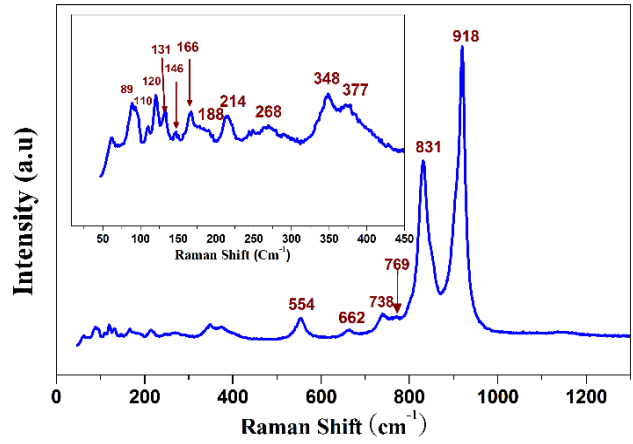

Figure 6. Room temperature Raman spectrum for AgCo$_2$V$_3$O$_{10}$. The inset shows the low frequency region 50-450 cm$^{-1}$

### 3.4. Magnetic properties

Figure 7 presents the molar magnetic susceptibility $\chi(T) = M(T)/H$ measured for the polycrystalline AgCo$_2$V$_3$O$_{10}$ sample 1 (S1)in the temperature interval 2 K $<T<$ 300 K for both ZFC and FCC regimes and the applied magnetic field $H = 100$ Oe. The inset shows the inverse molar magnetic susceptibility$\chi^{-1}(T)$ obtained in ZFC regime.

The data of Fig. 7 demonstrate: (1) the irreversible susceptibility ($\Delta\chi = \chi_{ZFC} - \chi_{FCC} \neq 0$) for $T<T_g \approx 5$ K and (2) the anomaly in the reversible $\chi(T)$ at $T_{N1} \approx 18$ K (see also Fig. 13).

At higher temperatures, the reversible molar magnetic susceptibility versus $T$ follows the Curie-Weiss law:

$$\chi(T) = \frac{C}{T+\theta} \qquad (1)$$

with the negative value of the Weiss constant $\theta = -18.88$ K characteristic of antiferromagnetic materials (see the inset in Fig. 7). The fitting gives Curie constant $C = 6.60$ emu K mol$^{-1}$ per formula unit which yields to $C = 3.30$ emu K mol$^{-1}$ per Co$^{2+}$. The calculated effective magnetic moment/Co$^{2+}$ was found $\mu_{eff} = 5.13\mu_B$. This value is in good agreement with that calculated considering spin−orbit coupling hypothesis for Co$^{2+}$(d$^7$, $S = 3/2$, $\mu_{LS} = 5.20\mu_B$) due to the unquenched orbital contribution commonly observed in Co$^{2+}$[41–43]. The $g$ factor, calculated from the Curie constant $(\mu_{eff})^2 = 8C = g^2 S(S+1)$ of 2.65 is similar to those reported for several cobalt-based compounds [44,45], indicating strong spin–orbital coupling and anisotropy for Co$^{2+}$.

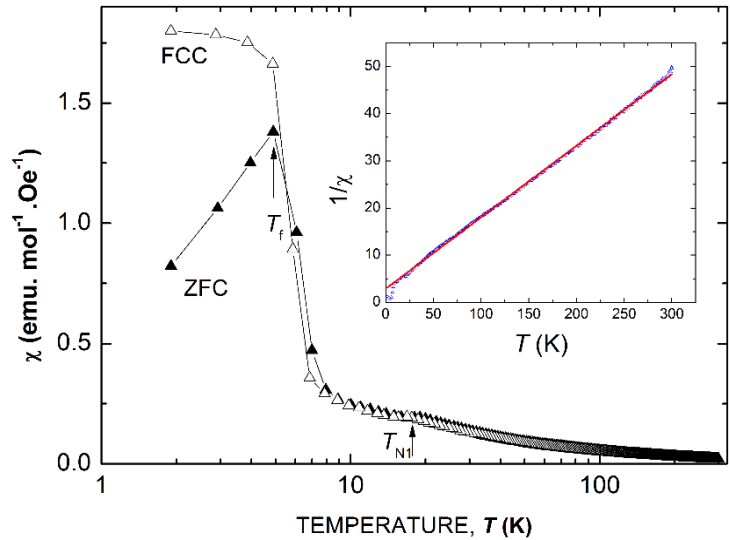

Figure 7. Magnetic susceptibility $\chi(T)$ measured for the polycrystalline for AgCo$_2$V$_3$O$_{10}$ in ZFC and FCC regimes under applied magnetic field $H = 100$ Oe. Arrows correspond to the field-dependent "spin glass" $T_f(H)$ and Neel $T_{N1}(H)$ temperatures. The inset presents the inverse ZFC magnetic susceptibility versus temperature; the solid red line is the Curie–Weiss law fit (see the text).

Figure 8 presents results of ZFC-FCC measurements of $\chi(T)$ for various applied magnetic fields. The results demonstrate a vanishing of the thermal hysteresis with the field increasing. A more detailed analysis, however, revealed the low temperature anomaly in $\chi(T)$ for all measured field, i. e. even in the reversible regime.

Figure 9 illustrates this fact, where we have plotted reversible $\chi(T)$ measured for $H = 25$ kOe and $H = 40$ kOe: two susceptibility peaks at $T_{N1} = 18$ K and $T_{N2} = 10$ K are clearly seen for $H = 25$ kOe. The peak at $T_{N1} = 17.6$ K and a "shoulder" at $T < T_{N1}$ [resulting from a superposition of two closely situated peaks in $\chi(T)$] can be also seen for $H = 40$ kOe.

The magnetic field - temperature ($H$-$T$) phase diagram in Fig.10 summarizes the obtained results.

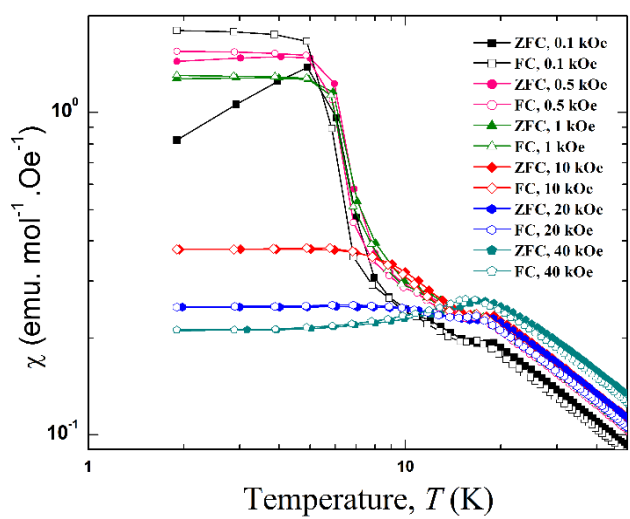

Figure 8. Low-temperature portions of $\chi(T)$ measured in ZFC and FCC regimes for different applied fields.

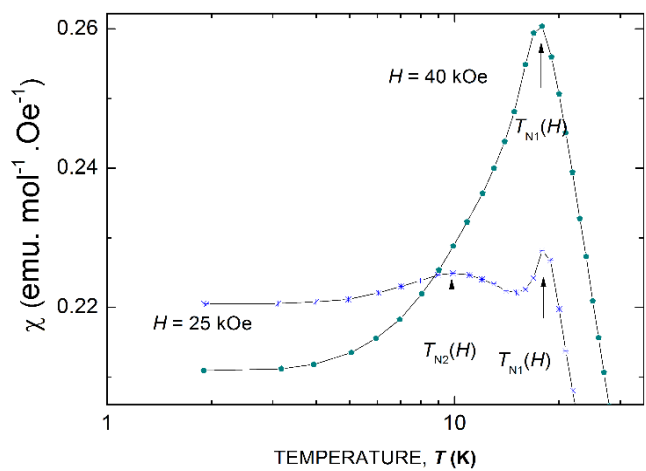

Fig. 9. Reversible magnetic susceptibility measured for $H = 25$ kOe and $H = 40$ kOe.

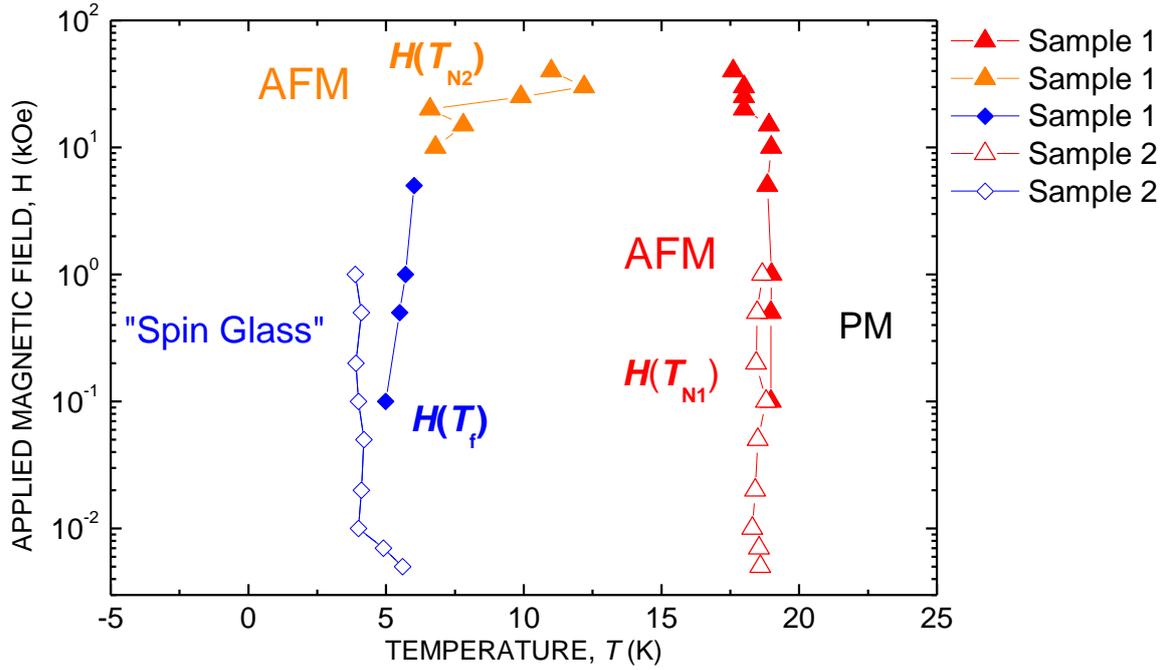

Fig. 10. Magnetic field - temperature phase diagram ($H$-$T$) constructed on the base of our experimental results for the samples S1 and S2; $H(T_f)$ data points are obtained as shown in Fig. 7; $H(T_{N1})$ and $H(T_{N2})$ correspond to two antiferromagnetic transitions [see Figs. 7 and 9; $T_{N2}(H)$ for $H$ = 30 kOe and $H$ = 40 kOe were found at minima in $d\chi(T)/dT$ occurring for $T<T_{N1}(H)$]; open symbols (◊) and (△) correspond to $H(T_f)$ and $H(T_{N1})$ obtained from the SQUID measurements for the sample 2.

A maximum in $\chi_{ZFC}(T)$ at $T_f$ ~5 K ($H$ = 100 Oe) as well as the difference between $\chi_{ZFC}(T)$ and $\chi_{FCC}(T)$ for $T<T_f$ (Fig. 7) resemble behavior of ordinary spin glasses (SG), see e. g., ref.[1].

However, two observations point out towards the unconventional character of the frozen spin state in $AgCo_2V_3O_{10}$: (1) contrary to conventional SG, $T_f(H)$ is field-independent, except of the very low fields (H < 10 Oe, S2) or even increases (S1) with the field increasing (Fig. 10); (2) Secondly, ac susceptibility measurements, see Fig. 11, revealed the absence of a frequency dependency of $T_f$, contrary to that expected for conventional SG [1].

Furthermore, the ferromagnetic (FM)-type $M(H)$ hysteresis loops measured for $T<T_f(H)$, see Fig. 12, indicate a coexisting of AFM and FM interactions in $AgCo_2V_3O_{10}$. The simultaneous occurrence of FM and AFM interactions implies their competition resulting in the spin-glass-like state.

In attempts to understand the above findings, we make a comparative analysis with other magnetically frustrated /spin-disordered systems. In fact, the "anomalous" increase of $T_f$ with $H$ (Fig. 10) has been known for antiferromagnetic spin glasses, where SG states develops at $T_f(H) < T_N(H)$ [46,47].

In contrast to the reentrant SG in ferromagnetic materials, where $T_f(H)$ is decreasing function of the field, in antiferromagnetic SG, applied magnetic field suppresses not only the glassy state but also antiferromagnetic correlations, resulting in the $T_f$ increase.

Assuming that $T_f(H)$ and $T_{N2}(H)$ in Fig. 10 is a single " spin frozen" phase boundary, the $H$-$T$ phase diagram resembles that predicted for frustrated AFM [46,48] where SG state takes below the AFM phase transition temperature.

However, our data obtained for $H>10$ kOe don´t show signatures of spin glasses. It seems, the applied magnetic field triggers AFM or some sort of intermediate (mixed) phase, as seen e. g. in geometrically frustrated magnets [5]. The oscillating character of $T_{N2}(H)$ (see Fig. 10) corroborates such a possibility pointing out on a competition between SG and AFM orders.

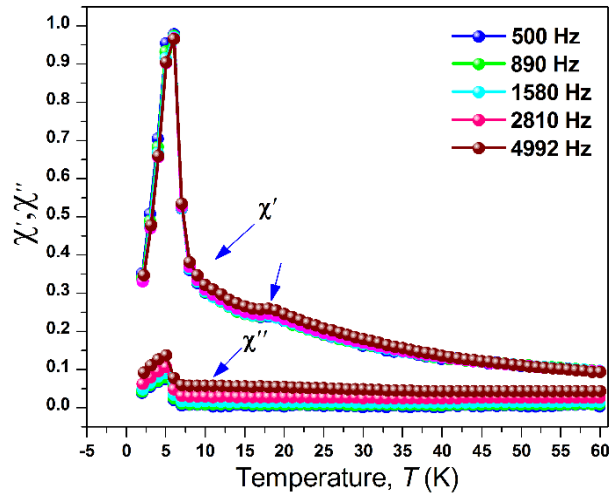

Figure 11. Temperature dependence of ac susceptibility, the real part $\chi'$ and the imaginary part $\chi''$ at different frequencies for AgCo$_2$V$_3$O$_{10}$ ($H_{ac}$= 10 Oe).

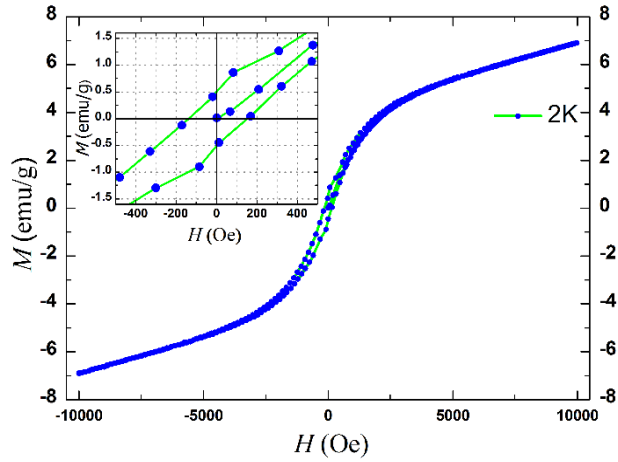

Figure 12. Ferromagnetic hysteresis loop $M(H)$ measured at $T = 2$ K. The inset shows a low-field portion of the loop.

Magnetization ZFC-FCC (susceptibility) measurements performed at low enough applied magnetic fields provide a further evidence for the coexisting AFM and FM correlations. As Figure 13 illustrates, at low enough magnetic fields and temperatures, $M_{ZFC}(T,H)$ is negative.

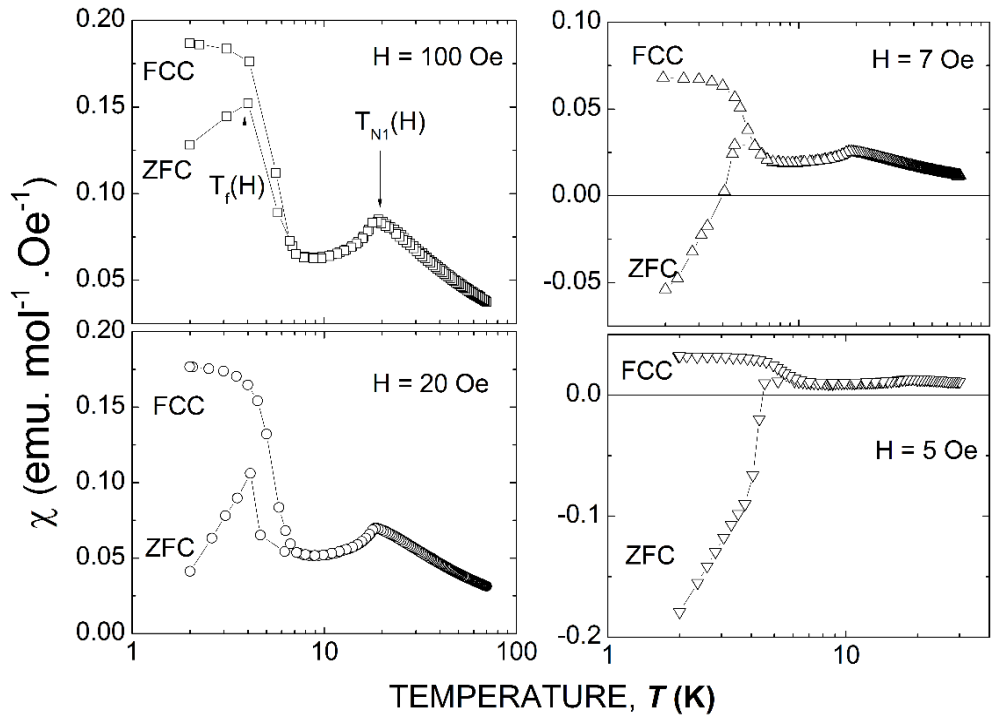

Fig. 13. SQUID magnetization (susceptibility) measured for the polycrystalline $AgCo_2V_3O_{10}$ (Sample 2) in ZFC and FCC regimes under various applied magnetic fields. Arrows correspond to the "spin glass" $T_f(H)$ and Neel $T_{N1}(H)$ temperatures.

This apparent diamagnetic response has been reported for various materials that possess both antiferromagnetic (AFM) and ferromagnetic (FM) components of spin ordering, see e. g., Refs.[6,49,50]. While the origin of the anomalous magnetization sign reversal is not yet clarified, its occurrence is essentially related to the presence of both AFM and FM interactions (for the review article see Ref. [51]).

It is particular interesting, in the context of the present study, note that (i) the anomalous magnetization sign reversal, (ii) the frequency independence of the « freezing » temperature $T_f$, and (iii) $T_f$ enhancement (or its independence) with the applied magnetic field observed for $AgCo_2V_3O_{10}$ have been also reported for the geometrically frustrated strongly spin-fluctuated $Cu_2Cl(OH)_3$[7].

For the completeness, we have performed the irreversible magnetization time relaxation measurements. Figure 14 presents the time variation of isothermal ZFC magnetization $M_{ZFC}(t)$ recorded in an applied magnetic field $H = 100$ Oe for $T = 2$ K. The observed logarithmic time variation of $M_{ZFC}(t)$ in a broad time window:

$$M(t) = M_0 + S \cdot \ln(1 + t/t_0) \qquad (2)$$

is the characteristic behavior of a variety of glassy systems, including spin glasses[1], where S is the magnetic viscosity coefficient, $t_0$ is a constant parameter which establishes a reference time and $M_0$ equals to the magnetization at the starting time of measurements.

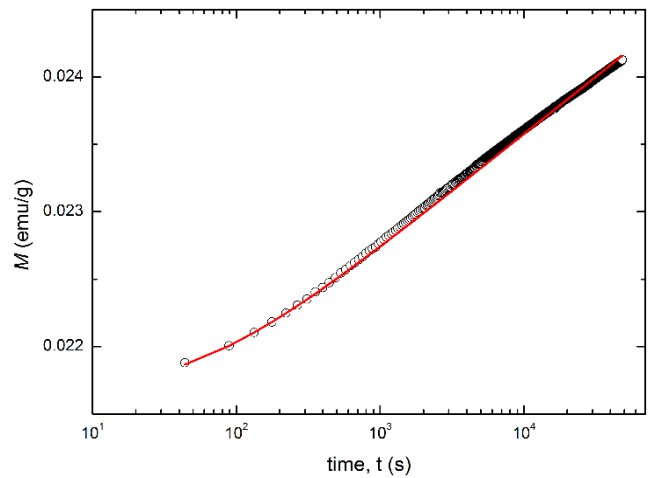

Fig. 14. ZFC magnetization $M_{ZFC}(t)$ time relaxation measured for $T = 2$ K , $T < T_f(H)$, and applied magnetic field $H = 100$ Oe. Red lineis obtained from Eq. (2) with $M_0(t = 44$ s$) = 2.17 \cdot 10^{-3}$ emu/g, $S = 3.7 \cdot 10^{-4}$ emu/g, $t_0 = 55$ s.

# 4. CONCLUSION

In conclusion, a new silver and cobalt-based vanadate, $AgCo_2V_3O_{10}$ was synthesized and characterized by single crystal X-ray diffraction, powder X-ray diffraction, Raman spectroscopy, and magnetic measurements. This new vanadate crystallizes in triclinic system with the centrosymmetric space group, $P\bar{1}$. Its structure consists of $Co_4O_{18}$ units and $V_3O_{10}$ groups forming through corner- sharing an open three-dimensional framework defining tunnels parallel to [001] where the $Ag^+$ cations are situated. DC and AC magnetic measurements revealed the presence of an unconventional spin-glass (SG) - type transition that increases with the applied magnetic field in addition to a frequency-independent behavior of the freezing temperature $T_f$. In addition, an open hysteresis loop was observed below $T_f$ which suggests the coexistence of ferromagnetic and antiferromagnetic correlations bringing the novel magnetic material, $AgCo_2V_3O_{10}$, into the class of magnetically frustrated SG systems.

**Supplementary material**

CCDC 1844136 contains the supplementary crystallographic data for this paper. The data can be obtained free of charge from The Cambridge Crystallographic Data Centre via [www.ccdc.cam.ac.uk/structures](www.ccdc.cam.ac.uk/structures).

- **Figure S1**. EDS spectra of single crystal and powder for $AgCo_2V_3O_{10}$.
- **Figure S2**. Le Bail refinement of the Powder X-ray diffraction pattern of $AgCo_2V_3O_{10}$. GOF = 1.27, $R_p$ = 7.72 and $R_{wp}$ = 9.72.
- **Table S1.** Fractional atomic coordinates and isotropic or equivalent isotropic displacement parameters ($Å^2$) of $AgCo_2V_3O_{10}$
- **Table S2.** Anisotropic displacement parameters ($Å^2$) of $AgCo_2V_3O_{10}$.
- **Table S3.** Selected bond lengths and angles of $AgCo_2V_3O_{10}$.

**Acknowledgements**

This work was done with the support of CNRST (Centre National pour la Recherche Scientifique et Technique) in the Excellence Research Scholarships Program. This work was also supported by the European H2020-MC-RISE-ENGIMA action (n° 778072). Yakov Kopelevich was supported by CNPq and FAPESP.